# Physics in the Machine: Integrating Physical Knowledge in Autonomous Phase-Mapping

**Authors:** A. Gilad Kusne[1,2]*, Austin McDannald[1], Brian DeCost[1], Corey Oses[4], Cormac Toher[4], Stefano Curtarolo[4], Apurva Mehta[3], Ichiro Takeuchi[2,5]

**Affiliations:**
[1] Materials Measurement Science Division, National Institute of Standards and Technology, Gaithersburg, MD 20899, US
[2] Materials Science and Engineering Department, University of Maryland, College Park, MD 20742, US
[3] Stanford Synchrotron Radiation Lightsource, SLAC National Accelerator Laboratory, Menlo Park, CA 94025, US
[4] Mechanical Engineering and Materials Science Department and Center for Autonomous Materials Design, Duke University, Durham, NC 27708, US
[5] Maryland Quantum Materials Center, University of Maryland, College Park, MD 20742, US

* Emails: aaron.kusne@nist.gov

## Abstract:
Application of artificial intelligence (AI), and more specifically machine learning, to the physical sciences has expanded significantly over the past decades. In particular, science-informed AI, also known as scientific AI or inductive bias AI, has grown from a focus on data analysis to now controlling experiment design, simulation, execution and analysis in closed-loop autonomous systems. The CAMEO (closed-loop autonomous materials exploration and optimization) algorithm employs scientific AI to address two tasks: learning a material system's composition-structure relationship and identifying materials compositions with optimal functional properties. By integrating these, accelerated materials screening across compositional phase diagrams was demonstrated, resulting in the discovery of a best-in-class phase change memory material. Key to this success is the ability to guide subsequent measurements to maximize knowledge of the composition-structure relationship, or phase map. In this work we investigate the benefits of incorporating varying levels of prior physical knowledge into CAMEO's autonomous phase-mapping. This includes the use of *ab-initio* phase boundary data from the AFLOW repositories, which has been shown to optimize CAMEO's search when used as a prior.

## Introduction
Machine learning (ML) application into the physical sciences poses interesting challenges of data sparsity, high data collection cost, high data complexity, and learning intricate functional relationships. Regarding data cost and sparsity, obtaining new data involves performing very complex, resource-intensive, and time-consuming experiments in the lab or *in silico*. Performing a successful experiment requires hours to months of expert time using equipment often costing hundreds of thousands to millions of dollars (e.g., microdiffraction at synchrotron beamlines). Additionally, the expertise needed is measured in years past doctorate graduation. As a result, many physical science ML challenges must learn from a small number of observations. Furthermore, obtaining the target information such as the stoichiometric composition of a material with optimal properties, may require mapping the relationship between numerous input parameters and target variables; i.e., the relationship between elemental composition and functional properties. With each new input parameter, the number of potential experiments grows exponentially. Consequently, the data obtained from costly experiments only sparsely represent a vast space of all possible experiments.

Confounding factors also include data complexity and the complexity of target relationships to be learned. Physical science data is often information rich. For instance, a Laue diffraction image from a material specimen contains information not only about crystal structures present in the sample, but also about their distribution, orientations, grain sizes, and crystallinity. Poor signal to noise and measurement setup-based signals, such as peaks due to the Cu K-β spectral profile which may vary from instrument to instrument, can overwhelm features of interest. As a result, combining data from multiple instruments and studies can be highly involved. Furthermore, the relationships investigated with this data tend to be complex. This is particularly true of many technologically relevant materials; for example, the relationship between a ferroelectric material's microstructure and its piezoelectric response.

These challenges are often not shared by non-science application domains in which common ML methods arose, such as deep learning. For these domains, semi-uniform data and labels can be collected rapidly and cheaply. For instance, labels for text and object images are freely provided by internet service users seeking to prove that they are not bots[1]. No specialized expertise or equipment is needed, and data collection occurs in seconds. As a result, big data velocities and volumes are possible. The range of possible data for these domains is also bounded; for example, text images are bound by language and handwriting, car navigation is bound to roads, and chess moves are bound by the rules of the game. Typically, the goal is to optimize safely within these bounds, while scientific studies seek to explore edge cases.

Despite the additional challenges, science has a key advantage relative to common application domains – there are hundreds of years of literature containing theory and heuristics for guiding research. Scientific artificial intelligence (AI) focuses on encoding these rules (i.e., inductive bias) into AI frameworks to ensure that analysis results and predictions obey the scientific rules, and are therefore physically meaningful[2]. Restricting the solution space may offer an additional benefit of increasing data analysis speed. Probabilistic scientific AI incorporates uncertainty quantification and propagation into the analysis to better inform scientific decision making.

Scientific AI offers significant benefits for autonomous physical research systems[3], where AI controls experiment design, simulation, execution, and analysis. For these systems, scientific AI can ensure that prior physical knowledge informs the selection of subsequent experiments, and that each experiment is selected to obtain maximal information. While scientific knowledge can be encoded at multiple levels of the autonomous AI pipeline—from data representation through the performance measure used to update model parameters—much of the reported successes use off-the-shelf machine learning methods. This includes active learning[4] algorithms—machine learning algorithms dedicated to optimal experiment design, which are used to determine each subsequent experiment to be performed. Applications of off-the-shelf active learning algorithms include the use of genetic optimization for carbon nanotube process optimization[5], Gaussian process upper confidence bounds to optimize molecular mixtures for photocatalysis[6], and estimate optimization for $CO_2$ electrocatalysis[7]. These successes of easily integrable, off-the-shelf active learning create opportunities and physical platforms where scientific AI may provide even greater research acceleration.

Recent work by Kusne and coworkers[8] demonstrates an autonomous physical research system for accelerating composition-phase-mapping and materials optimization, specifically the identification of optimal compositions that maximize some desired properties within a targeted search space. The autonomous system is driven by CAMEO (closed-loop autonomous materials exploration and optimization). This scientific AI algorithm was placed in control of the Stanford Synchrotron Radiation Lightsource high-throughput diffraction system, guiding each subsequent x-ray diffraction experiment, resulting in the discovery of a best-in-class phase change memory material. CAMEO was shown to accelerate materials optimization compared to standard methods by exploiting the materials composition-structure-property relationship to guide subsequent experiments. Toward this goal, CAMEO performs active phase-mapping – investigating subsequent compositions that provide maximal knowledge of the composition-structure relationship as represented by the composition-phase map. The structural phase map is fundamental to materials optimization as functional property extrema tend to occur within specific phase regions (e.g., magnetism and superconductivity) or along phase boundaries (e.g., martensitic transformation and morphotropic phase-boundary piezoelectrics). Knowledge of the phase map is used to guide materials optimization toward more promising regions of the search space.

Active phase-mapping can be thought of as an exploratory task to learn the composition-structure relationship. The composition space is segmented into regions based on which phases are present. To improve the performance of active phase-mapping, multiple levels of scientific knowledge can be incorporated, including density functional theory (DFT) data from the AFLOW.org repositories[9,10]. This work investigates the impact of varying levels of incorporated physical knowledge on active phase-mapping performance. A full list of the algorithms studied, their varying levels of incorporated physical knowledge, and how the physical knowledge is encoded is provided in the Methods Table M1. Performance is explored for the benchmark ternary materials system of Fe-Ga-Pd[11].

## Discussion

For this study, the level of scientific information in the active phase-mapping algorithm is varied by two factors – the first being the phase-mapping method. The structural phase-mapping method consists of 1) identifying the composition-phase map for samples with measured composition and x-ray diffraction patterns and then 2) extrapolating to samples without measured diffraction. Two phase-mapping methods are investigated. The first method uses go-to, off-the-shelf ML methods for clustering and classification: agglomerative hierarchical cluster analysis (HCA) with a cosine dissimilarity measure applied to the diffraction patterns[12] and a first-nearest neighbor algorithm for extrapolating phase region labels across the composition space. The alternative method uses the scientific AI phase-mapping method of CAMEO. The CAMEO phase-mapping method employs a Bayesian graph-based algorithm to identify the probability of each composition sample belonging to each structural phase region. As a result, this method can generate a list of structural phase diagrams and their likelihoods. The method selects the most likely phase diagram based on the given data.

The optimal experiment design (OED) algorithm is the second factor varied, determining the sequence of samples to measure for diffraction data. Four methods are employed, as list in the column "Active Learning Sampling Method" in Table 1. The first method measures samples sequentially by their composition spread index [see Supplementary Figure 4(b) of Ref. [8]]. The next method selects samples randomly using a uniform distribution over composition – a common exploratory active learning benchmark when the goal is gaining global knowledge of a search space. The third method selects each subsequent sample so that it minimizes total expected phase region misclassification error, here described as risk minimization[8]. This method was shown to target subsequent measurements along uncertain portions of the structural phase boundaries. The used risk minimization method requires a graph-based data representation and as such can only be combined with the graph-based CAMEO phase-mapping method. The sequential, random, and risk minimization methods are also compared to the performance of selecting 10 % of the composition spread samples that provide good composition space coverage [see Supplementary Figure 4(a) of Ref. [8]]. The 10 % coverage method is expected to provide good exploratory sampling and provide similar performance to the uniform random sampling as averaged over many runs.

As an additional modality for introducing prior physical knowledge, a Bayesian probabilistic prior over the phase map is implemented. The prior is derived from DFT calculations for the bulk Fe-Ga-Pd phase diagram as calculated by AFLOW[9,10], with phase boundary data resolved by the AFLOW-CHULL[13] module (see Supplementary Figure 2 of Ref. [8]). The probabilistic prior is graph-based, defining the probability of materials belonging to the same phase region, and as such is demonstrated only in combination with the graph-based CAMEO phase-mapping method and the risk minimization OED method.

Autonomous phase-mapping performance is shown in Figure 1(a) using the modified Fowlkes-Mallow Index (FMI) performance measure[8], comparing the machine learning based phase-mapping results with expert labeled results. Here performance is averaged over 100 runs with the plot indicating the average performance with 95 % confidence intervals (except for the 10 % coverage OED method). Each autonomous phase-mapping method is indexed and described in Table 1. The index number corresponds to a rank of performance at iteration 27, where 10 % of the samples have been measured, allowing for comparison with the 10 % sampling method. This is also the earliest iteration at which CAMEO Method 8 achieves an average performance of 85 %.

In investigating the relative performance, it is interesting to note that the methods first group by OED method and then by phase-mapping method. For each OED method, the more physics-informed CAMEO phase-mapping method out-performs the off-the-shelf alternative. A complicating factor is that the off-the-shelf method is limited to phase-mapping with 5 structural phase regions, while the CAMEO phase map method allows the number of phase regions to vary and converge to an optimal. To ensure that the increase in performance is not due to an increase in the number of phase regions, i.e., model complexity, the average number of phase regions over the 100 runs is provided in Figure 1b.

OED performance also increases with greater prior physical knowledge. While sequential OED (Methods 1 and 2) simply contains information of sample location on the wafer, the use of the random and 10 % sampling OED (Methods 3 through 6) assume that greater coverage of the composition space will provide more phase map knowledge. Finally, risk minimization (Methods 7 and 8) provides the best performance, building on the assumption that the most informative samples lie along phase boundaries.

Of particular interest is the fact that introducing prior information from AFLOW of the Fe-Ga-Pd bulk DFT phase diagram calculation (Method 8) achieves superior performance at lower iterations and then converges to a performance beneath those achieved by other methods including the CAMEO Method 7. Initially, when few diffraction patterns have been measured, the strong prior provides a correcting bias. However, as more data is obtained, the DFT-based bias pulls away from the correct answer for the thin film composition phase map.

For active phase-mapping, an increasing amount of physics information incorporated in the scientific ML provides better performance. While this improvement is demonstrated for a 2-dimensional composition space (3-simplex), it is expected that improvements will be more significant when searching higher dimensional spaces, as structural phase boundaries become exponentially sparser with increasing number of dimensions[13,14]. Similarly, the search for optimal materials becomes increasingly difficult. As a result, the use of physics-informed active phase-mapping --- through a combination of experiments and ab-initio calculations --- is expected to become ever more important in guiding the search for novel, advanced materials.

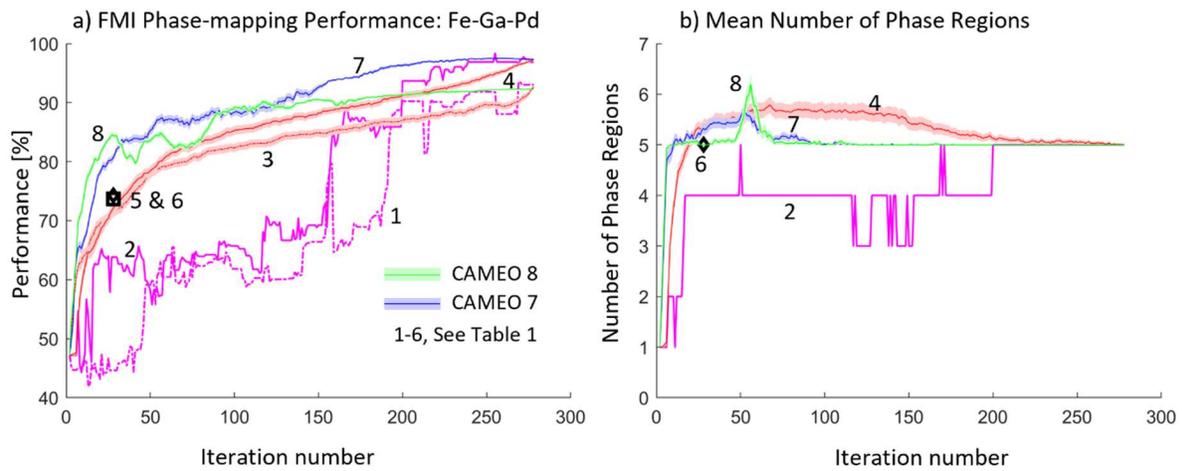

*Figure 1. Fowlkes Mallow Index (FMI) phase-mapping performance for Fe-Ga-Pd material system. a) FMI performance for the set of methods listed in Table 1. The average performance is indicated with 95 % confidence intervals for methods 5 through 8 which were run 100 times with uniform randomly selected initial sample composition, b) the average number of phase regions (over the 100 iterations) used in phase-mapping for the CAMEO phase-mapping methods. If the method does not appear here, the number of phase regions was fixed at 5.*

*Table 1. Phase-mapping methods in order of performance (descending) at iteration 27.*

| Algorithm Index | Phase-mapping Method | Prior | Active Learning Sampling Method | Mean FMI Performance for iteration 27 [%] |
|---|---|---|---|---|
| 8, 'CAMEO' | CAMEO Phase-mapping | Y | Risk Minimization | 85 |
| 7, 'CAMEO' | CAMEO Phase-mapping | N | Risk Minimization | 80 |
| 6 | CAMEO Phase-mapping | N | 10 % | 74 |
| 5 | HCA + 1NN | N | 10 % | 74 |
| 4 | CAMEO Phase-mapping | N | Random | 72 |
| 3 | HCA + 1NN | N | Random | 71 |
| 2 | CAMEO Phase-mapping | N | Sequence | 64 |
| 1 | HCA + 1NN | N | Sequence | 45 |

HCA = Hierarchical Cluster Analysis
1NN = 1-Nearest Neighbor

# Methods
## M1: Scientific AI

Table M1. Scientific AI physical knowledge and encoding method.

| Algorithm | Physical knowledge | Encoding Method |
|---|---|---|
| **Data Analysis** | | |
| HCA | Diffraction similarity identified by peak location rather than intensity. | Use of Cosine dissimilarity measure[12] |
| CAMEO Phase-mapping[8] | Phase regions are contiguous and phase boundaries are continuous | 1. If two or more sets of vertices share the same phase region label but are not connected by vertex neighbors, differing labels are assigned to the disconnected sets. 2. The Markov Random Field smoothness constraint[15] |
| | Materials of similar synthesis and processing parameters have similar properties | 1. Markov Random Field smoothness constraint[15] 2. Harmonic Energy Minimization for label propagation[16] |
| | Abundances of phases is non-negative | Karush–Kuhn–Tucker conditions[17] |
| | X-ray diffraction intensity is non-negative | Karush–Kuhn–Tucker conditions[17] |
| | Soft Gibbs Phase Rule - Upper bound limit on number of constituent phases | Upper limit on number of endmember limits allowed in each phase region |
| | Identified endmembers should be physically realizable | Volume constraint on identified / predicted endmembers |
| Phase-mapping Prior | DFT phase map is predictive of bulk phase diagram. Structure is a good predictor of functional property and vice versa | Bayesian prior through similarity kernel For more information see Ref [8,13] M1c Phase Mapping: Phase mapping prior. |
| **Knowledge Propagation** | | |
| 1-NN | Samples of similar composition are likely to have similar phase. | As more samples are measured, the distance between samples in composition space gets smaller, so neighbors are more likely to have similar structure. |
| HEM | Phase regions are cohesive. Quantified likelihood for each sample belonging to each phase region due proximity in composition | Graph representation of composition space. Label propagation through graph. Labels uncertainty propagation. |
| **Active Learning** | | |
| Sequence | None | |
| 10 % Sampling | Samples chosen to be well distributed in composition space | Samples evenly distributed across composition space. |
| Uniform Random Sampling | Sampling uniformly will give general coverage of the composition space. | |
| Risk Minimization | Each sample quantified for its potential impact on improving total phase map performance. Targets phase boundaries. | Minimize total phase region misclassification probability for the entire phase map. |

## M2 Statistics and Performance Metrics
### Confidence Interval

The 95 % confidence interval was computed for the variable of interest over 100 experiments at the given iteration with:

$$CI_{95} = \left(\frac{\sigma}{\sqrt{n}}\right) F^{-1}(p, v) \qquad (12)$$

Where $F^{-1}$ is the inverse of the Student's t cumulative distribution function, $\sigma$ is the standard deviation, $n = 100$ is the number of experiments, $p = \{2.5\,\%, 97.5\,\%\}$, and $v = 99$ is the degrees of freedom.

## Phase-mapping Performance

Phase-mapping performance is evaluated by comparing phase region labels determined by experts with those estimated by CAMEO for the entire phase map (after the knowledge propagation step). To evaluate system performance, the Fowlkes-Mallows Index (FMI) is used, which compares two sets of cluster labels. The equations are presented below for the expert labels $l \in L$ and the ML estimated labels $\hat{l} \in \hat{L}$, where the labels are enumerated $L \to \mathbb{N}$ and $\hat{L} \to \mathbb{N}$.

If the number of phase regions is taken to be too large by either the user or the ML algorithm while the phase-mapping is correct, some phase regions will be segmented into sub-regions with the dominant phase boundaries preserved. For example, peak shifting can induce phase region segmentation[44]. To ensure that the performance measures ignore such sub-region segmentation, each estimated phase region is assigned to the expert labeled phase region that shares the greatest number of samples. The number of phase regions is monitored to ensure that increases in model accuracy are not driven by increases in model complexity.

$$\text{Fowlkes-Mallows Index: } FMI = TP/\sqrt{(TP + FP)(TP + FN)} \quad (13)$$

$$TP = \frac{1}{2} \sum_i \sum_j (l_i = l_j \,\&\, \hat{l}_i = \hat{l}_j) \quad (14)$$

$$FP = \frac{1}{2} \sum_i \sum_j (l_i \neq l_j \,\&\, \hat{l}_i = \hat{l}_j) \quad (15)$$

$$FN = \frac{1}{2} \sum_i \sum_j (l_i = l_j \,\&\, \hat{l}_i \neq \hat{l}_j) \quad (16)$$

## M3. Implementation

The methods were implemented in MATLAB*. Built-in functions were used for agglomerative hierarchical cluster analysis and 1-nearest neighbors.

**NIST Disclaimer:** Certain commercial equipment, instruments, or materials are identified in this report in order to specify the experimental procedure adequately. Such identification is not intended to imply recommendation or endorsement by the National Institute of Standards and Technology, nor is it intended to imply that the materials or equipment identified are necessarily the best available for the purpose.

**Competing interests:** The authors declare no competing interests.

**Data Availability**
Data that supports the findings of this study have been deposited in the github repository and can be found with the following github or DOI link:
https://github.com/KusneNIST/CAMEO_NComm
https://doi.org/10.5281/zenodo.3998287

**Code Availability**
The code can be found at the following github repository or using the following DOI link:
https://github.com/KusneNIST/CAMEO_NComm
https://doi.org/10.5281/zenodo.3998287

Acknowledgements

The authors thank Xiomara Campilongo and Marco Esters for fruitful discussions.

Contributions:
AGK performed the computations and analysis with input from IT, AM, AM, and BD. SC, CO, and CT provided the density functional theory data for the computations. The authors wrote the text together.